\documentclass[aps,prb,twocolumn,preprintnumbers,amsmath,amssymb,superscriptaddress,nofootinbib]{revtex4-2}
\usepackage{graphicx}
\usepackage{dcolumn}
\usepackage{bm}
\usepackage{color}
\usepackage[usenames,dvipsnames]{xcolor}
\usepackage{hyperref}
\usepackage{changes}

\hypersetup{
    bookmarks=false,        
    colorlinks=true,        
    linkcolor=blue,        
    citecolor=blue,        
    filecolor=blue,      
    urlcolor=blue
}

\newcommand{\be}{\begin{equation}}
\newcommand{\ee}{\end{equation}}
\newcommand{\bea}{\begin{eqnarray}}
\newcommand{\eea}{\end{eqnarray}}
\newcommand{\eq}[1]{Eq.~(\ref{#1})}
\newcommand{\Sec}[1]{Sec.~\ref{sec:#1}}
\newcommand{\fig}[1]{Fig.~\ref{Fig:#1}}

\newcommand{\e}{\varepsilon}
\newcommand{\w}{\omega}
\newcommand{\s}{\sigma}

\newcommand{\up}{\uparrow}
\newcommand{\down}{\downarrow}

\newcommand{\dd}{\partial}

\newcommand{\ie}{\textit{i.e.}}
\newcommand{\eg}{\textit{e.g.}}

\newcommand{\op}[1]{\hat{#1}}
\newcommand{\dk}{\op{d}^\dagger}

\newcommand{\ck}{\op{c}^\dagger}
\newcommand{\da}{\op{d}^{}}

\newcommand{\ca}{\op{c}^{}}

\newcommand{\vk}{\mathbf{k}}
\renewcommand{\s}{\sigma}
\renewcommand{\up}{\uparrow}
\renewcommand{\down}{\downarrow}
\newcommand{\hc}{h.c.}

\newcommand{\avg}[1]{\left\langle #1 \right\rangle }
\newcommand{\dB}{\frac{\dd}{\dd B}}

\begin{document}

\title{Majorana coupling and Kondo screening of localized spins}

\author{Krzysztof P.~W\'{o}jcik}
\email{krzysztof.wojcik@mail.umcs.pl}
\affiliation{Institute of Physics, Maria Curie-Sk\l{}odowska University, 20-031 Lublin, Poland}
\affiliation{Institute of Molecular Physics, Polish Academy of Sciences, ul.~Smoluchowskiego 17, 60-179 Pozna{\'n}, Poland}

\author{Piotr Majek}
\affiliation{ISQI, 
	Faculty of Physics, Adam Mickiewicz University, 
	61-614 Pozna{\'n}, Poland}

\date{\today}

\begin{abstract}

We perform a theoretical analysis of the fate of local magnetic moment 
of a quantum dot coupled to a normal metallic lead and a topological superconducting 
wire hosting Majorana modes at the ends. 
By means of simple analytical tools and numerical renormalization group calculations 
we show that the proximity of Majorana mode reduces the magnetic moment from $1/4$,
characteristic of a free spin $1/2$, to $1/16$. The coupling to the normal lead then 
causes the Kondo effect, such that the magnetic moment is fully screened below the 
Kondo temperature. The latter is vastly increased for strong coupling to Majorana 
mode.

\end{abstract}

\maketitle

\section{Introduction}
\label{sec:intro}

The quest for realization of Majorana modes (MMs) in solid state 
is motivated mainly by applications \cite{Kitaev2003Jan,Sau2010Jan,Clarke2011Jul} 
and promising experimental results 
\cite{Mourik2012May}.
A hope for fault-tolerant quantum computation using MMs stems from 
their topological protection against local disturbances \cite{BraidingReview}.
As a consequence, a rich field of research focused on realizing and 
manipulating MMs emerged, as summarized in a number of reviews
\cite{Sato2017May,Prada2020Rev,Lutchyn2018May,Laubscher2021Aug}.

Recently, the field experiences criticism, as summarized in Ref.~\cite{Frolov2021Apr},
concerning hasty publications with exaggerated conclusions. 
Furthermore, practical implementation for useful computation would 
require large number of Majorana devices, which does not seem feasible 
soon.
Here we leave the mainstream application-oriented approach, and address 
basic theoretical questions concerning interplay of MMs with strongly 
interacting mesoscopic systems, hoping to gain some understanding.
This relatively unexplored direction has been pioneered by Ref.~\cite{Cheng2014Sep},
and remains relevant especially in the context of transport properties of 
quantum dots proximitized by topological superconducting wires 
\cite{Lee2013Jun,Ruiz-Tijerina2015Mar,Weymann2017Apr,Vernek2019,Majek2022Feb}. 
We focus on the interplay between the Kondo effect \cite{Hewson1997}, 
and the local MM-spin coupling, studying the minimal model, 
as elaborated further. 
We analyze it by looking at the fate of magnetic moment localized on a quantum dot
at low temperatures.

\section{Model and methods}
\label{sec:model}

\begin{figure}[tb]
	\centering
	\includegraphics[width=0.5\columnwidth]{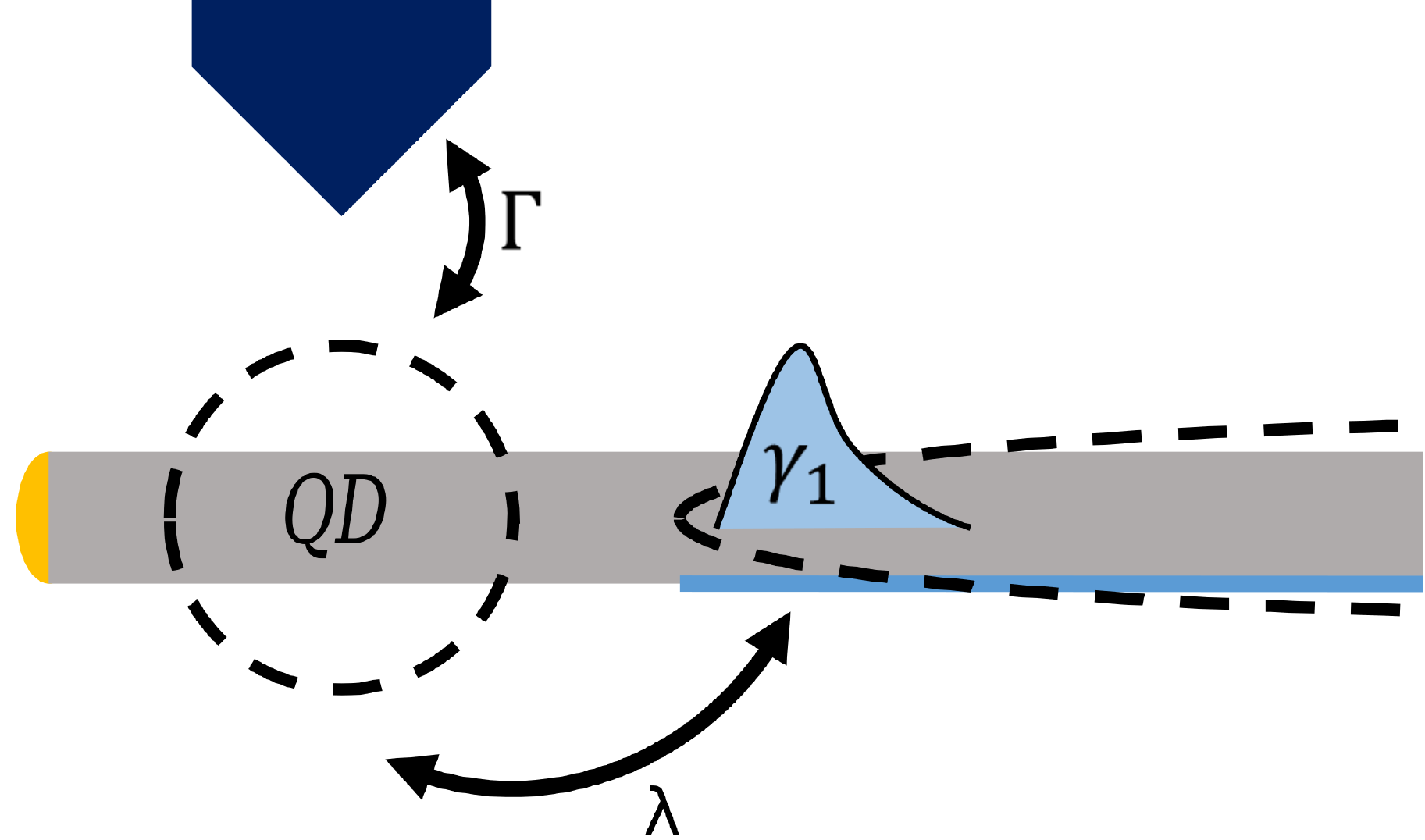}
	\caption{The scheme of the structure comprising a single quantum dot
	hybrydized with normal-metal contact with overall coupling strength 
	$\Gamma$, and coupled to long topological superconductor nanowire 
	hosting Majorana modes $\gamma_1$ and $\gamma_2$.
	}
	\label{Fig:scheme}
\end{figure}

We consider a single quantum dot (QD) coupled to one-dimensional topological 
superconducting nanowire, called further the Majorana wire (MW); see \fig{scheme}. 
MW is characterized by a superconducting gap, and a pair of 
(Majorana) modes at Fermi energy, which are strongly localized at the ends of 
the wire. 
Therefore, especially at temperatures $T$ much smaller than the superconducting 
gap, the only relevant coupling between QD and MW is the hopping of electrons
into and from these end-modes, and the in-gap Hamiltonian of the QD and MW can be 
written in the form \cite{Cheng2014Sep}
\be
H_{\rm DM} = \sum_\s \e\, \op{n}_\s + U \op{n}_\up \op{n}_\down
			+ \lambda (\dk_\down \op\gamma_1 + \hc) + i\e_M \op\gamma_1\op\gamma_2,
\label{H}
\ee
where 
$\e$ is QD energy level, 
$U$ -- the on-dot Coulomb repulsion, 
while $\lambda$ measures the QD-MW coupling strength.
The operators $\dk_{\s}$ creates spin-$\s$ electron at QD,
$\op{n}_\s = \dk_{\s}\da_{\s}$, 
and the Majorana operators $\op\gamma_1$, $\op\gamma_2$
are normalized such that $\{\op\gamma_1,\op\gamma_2\} = 1$.
The last term in $H_{\rm DM}$, proportional to $\e_{\rm M}$,
corresponds to overlap between Majorana modes, 
which is exponentially small for long wires 
and shall be neglected henceforth, \ie{} $\e_M=0$.
Note that only $\s=\;\down$ electrons are coupled to MW. 

Eq.~(\ref{H}), may at first glance be seen as a fusion of 
the Anderson-like impurity with the Kitaev chain model \cite{Kitaev2001},
where bulk states are completely removed and only in-gap states remain.
One of them is then coupled via hopping with QD.
In experiment the MW is a complex hybrid nanostructure, typically
comprising a semiconductor with strong spin-orbit coupling and
a conventional superconductor, almost fully spin-polarized with
the help of magnetic field. While such system is vastly more
complicated than the Kitaev model, it eventually leads to practically
spin-polarized $p$-wave superconductor, with energy gap and in-gap MMs.
These are the generic features we model simply by Eq.~(\ref{H}),
which serves very well at temperatures below the energy gap in the 
superconductor, as long as the MW is long enough for the coupling 
to its other side and $\e_{M}$ both to be neglected \cite{Prada2017Aug}.

QD is further attached to metallic electrode, modeled as non-interacting, 
with energy dispersion $\e_{\vk}$. This means the lead
part of the Hamiltonian is 
$H_{\rm L} = \sum_{\vk\s} \e_{\vk} \op{n}_{\vk\s}$, 
with $\op{n}_{\vk\s} = \ck_{\vk\s}\ca_{\vk\s}$ and 
$\ck_{\vk\s}$ creating corresponding electron. 
Finally, the hybridization term takes a form 
$H_{\rm h} = \sum_{\vk\s} v (\dk_{\s} \ca_{\vk\s} +\hc)$,
where $v$ is the tunneling matrix element and $\hc$ stands for 
Hermitian-conjugate term. The total Hamiltonian of the system 
is $H = H_{\rm DM} + H_{\rm h} + H_{\rm L}$.
In the calculations we take the wide-band limit, that is we assume 
the (normalized) density of leads states $\rho(\w)$ is a constant 
within a cut-off window, $\w \in [-D,D]$, and vanishes outside 
(we use $D=2U$). $\Gamma=\pi \rho(0) v^2$ measures the coupling
strength to normal lead.

In general, magnetic susceptibility $\chi$ is defined as a linear 
coefficient of response of spin polarization, induced by applying 
small external magnetic field $B$. 
In impurity or QD systems the relevant quantity is the impurity 
contribution to $\chi(T)$ \cite{Hewson1997}, 
which can be defined as
\be
\chi(T) = -\left. \dB \avg{\op{S}_z}_T \right|_{B=0} ,
\label{chi}
\ee
where $\op{S}_z$ is the $z$-th component of the QD spin,
$B$ is the magnetic field acting locally at QD
and $\avg{\ldots}_T$ denotes thermal expectation value.
The magnetic moment is simply $\mu(T) = T \chi(T)$.
To calculate these quantities numerically at given $T$, we add 
a small field $B\ll T$ into $H$, $H \mapsto H + g \mu_B B \op{S}_z$,
with gyromagnetic ratio $g$, Bohr magneton $\mu_B$, units of $B$ 
adjusted such that $g \mu_B = 1$, and $\op{S}_z = (\op{n}_\up - \op{n}_\down)/2$. Then, 
$\chi(T) = -\avg{\op{S}_z}_T/B$ follows from \eq{chi}.

To reliably solve the model in the Kondo regime we use
numerical renormalization group (NRG) technique \cite{Bulla2008}.
Our implementation is based on open-access code \cite{fnrg},
exploiting symmetries of charge parity and total spin-$\up$ 
electron number conservation.
We use discretization parameter $\Lambda=3$ and keep around
$N=1000$ states during NRG iteration.
We also provide a number of exact analytical results where possible.

\section{Results}
\label{sec:results}

For a free spin $S$ at low $T$ the magnetic susceptibility 
can be calculated directly as defined in \eq{chi},
\be
\chi_S(T) = \frac{S(S+1)}{3T}, 
\label{chiS}
\ee
which implies $\mu(T) = 1/4$ for $S=1/2$. This high fragility 
to magnetic field leads to vulnerability of the localized 
spins, often suppressed at low $T$ due to one of the following 
circumstances:
\begin{enumerate}
\item Ordering tendencies: 
	in practice spins are often coupled by exchange interaction $J$, 
	which in case of just $2$ local moments leave them in a singlet 
	[for antiferromagnetic (AFM) $J>0$] or triplet [for ferromagnetic 
	(FM) $J<0$] ground state at $T=0$, with $\avg{\op{S}_z}$ independent 
	of $B$ unless it exceeds the binding energy $\sim |J|$.
	The same mechanism leads to magnetic instabilities in the lattices
	possessing local moments, which tend to form a magnetic order 
	(FM or AFM, depending on the sign of effective exchange coupling 
	between the localized spins);
\item The Kondo effect: 
	the coupling to continuous bath drives the Kondo effect. Then, 
	a portion of free conduction electrons bind into a singlet with 
	a localized QD spin at $T$ below so-called Kondo temperature, $T_K$. 
	Then, the divergence of $\chi$ is suppressed below $T_K$, such 
	that for $T\ll T_K$ we have $\chi \sim 1/T_K$ and $\mu = 0$ 
	\cite{Hewson1997} (see in particular Appendix K there). 
\end{enumerate}
These two mechanisms typically compete with each other \cite{Doniach},
driving many strongly-correlated phases of matter 
\cite{Paschen2021Jan,Wojcik2021Jun}.

\begin{figure}[tb!]
	\centering
	\includegraphics[width=0.9\columnwidth]{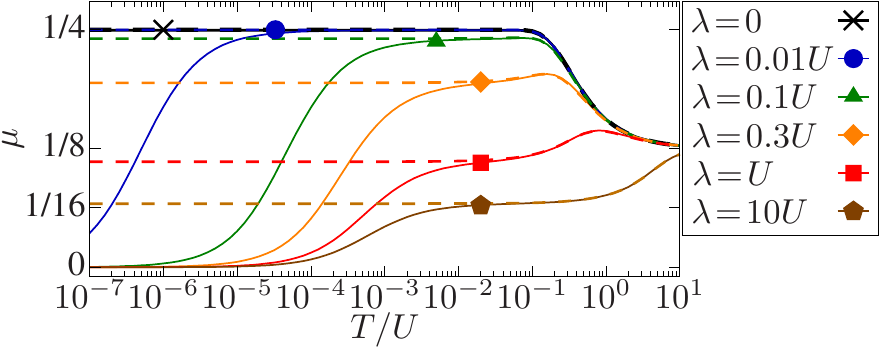}
	\caption{%
		Temperature dependence of magnetic moment $\mu(T)$
		for $\e=-U/2$ and $\Gamma=0.001U$ (solid lines, NRG results), 
		as well as $\Gamma=0$ [dashed lines, exact \eq{muDM}], 
		and different values of $\lambda$, as indicated in the figure.
	}
	\label{Fig:T}
\end{figure}
\begin{figure}[tb!]
	\centering
	\includegraphics[width=0.9\columnwidth]{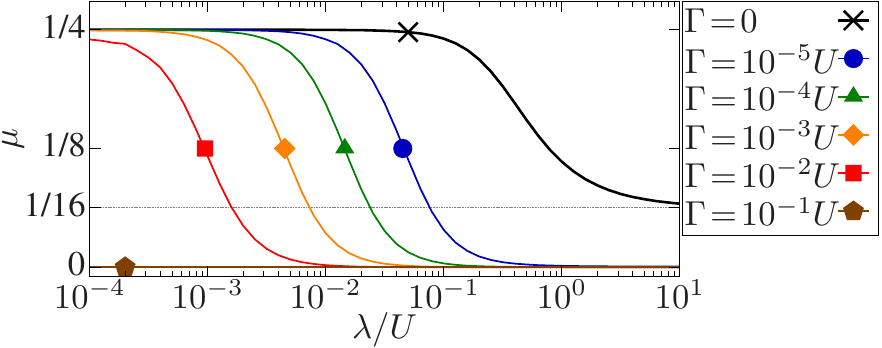}
	\caption{%
		The magnetic moment $\mu$ for $T=10^{-7}$ as a function of QD-MW coupling 
		$\lambda$ for indicated values of $\Gamma$.
	}
	\label{Fig:Maj}
\end{figure}

In the model introduced in \Sec{model}, the free spin behavior is present 
in the absence of QD--leads couplings, $\Gamma=\lambda=0$. A direct calculation
following from \eq{chi} gives then
\be
\mu_{\rm QD} = \frac{1}{4} \left[1 +\exp\left(-\frac{U}{2T}\right) \cosh\!\left(\frac{\delta}{T}\right)\right]^{-1} ,
\label{muQD}
\ee
where $\delta=\e+U/2$. 
In the limit $T/U \to 0$ this result asymptotically reaches $(4T)^{-1}$
expected from \eq{chiS} for $S=1/2$,
as long as $|\delta|$ does not exceed $U/2$ (QD is singly occupied then). 
This is shown (for $\delta=0$) in \fig{T} as the dashed black curve.

\eq{chi} can be exactly solved in a relatively simple form also 
for $\lambda\neq 0$, $\delta=0$ case, since $H_{\rm DM}$ can be 
diagonalized exactly \cite{Weymann2017Apr}. The result is
\be
\mu_{\rm DM} \!= \frac{1}{4} \! 
						\left[ \frac{w + 16u^2 T/U}{2 w^{3/2}} \!\tanh\!\left(\!\frac{\sqrt{w}}{4 T/U}\!\right)
						+\frac{1+4 u^2}{2 w}\right] ,
\label{muDM}
\ee
where we set $u=\lambda/U$ and $w=1+8u^2$. This is presented for 
several values of $\lambda$ as dashed lines in \fig{T}. 
As expected, for $\lambda=0$ we have $u=0$, $w=1$, and at low $T$ 
(when the argument of $\tanh$ function becomes very large) $\mu_{\rm DM} \to 1/4$.
However, for $\lambda>0$ the $T=0$ magnetic moment becomes 
\be
\mu_{\rm DM}^{T\to 0} = \frac{1}{4} \cdot \frac{1 + \sqrt{1+8u^2} + 4 u^2}{2 (1+8u^2)},
\label{muDM0}
\ee
which is plotted in \fig{Maj} with a thick black line.
Strikingly, $1/4 \geq \mu_{\rm DM}^{T\to 0} > 1/16$, 
that is the magnetic moment is \textit{partially suppressed} 
in the vicinity of MW. This reflects fractional nature of MMs.
Note that the suppressed fraction is not universal, 
and even the minimal value $\min(\mu_{\rm DM}^{T\to 0}) = 1/16$ 
does not correspond to any intuitive effective free spin $S$
in \eq{chiS}. Inverting it with $\chi_S=(16T)^{-1}$ gives
$S=(\sqrt{7}-2)/4\approx 0.161$.

The significance of this result can be better understood by
contrasting it with a result obtained for a QD proximitized by 
a conventional superconductor. Indeed, taking the 
BCS Hamiltonian with the states outside the gap integrated 
out \cite{Rozhkov2000Sep},
$
H_{\rm DS} = \sum_\s \e\, \op{n}_\s + U \op{n}_\up \op{n}_\down
			+ \Gamma_{\rm S} (\dk_\up \dk_\down + \hc),
$
instead of $H_{\rm DM}$ of \eq{H}, we get 
\be
\mu_{\rm DS} =  \frac{1}{4} \left[ 
				1+ 
				\cosh\!\left(\!\frac{\sqrt{\Gamma_{\rm S}^2+\delta ^2}}{T}\right) 
				\exp\!\left(\!-\frac{U}{2T}\right)
				\right]^{-1} \!\!\!\! .
\label{muSC}
\ee
This differs from \eq{muQD} only by replacement 
$\delta \mapsto \sqrt{\delta^2+\Gamma_{\rm S}^2}$,
which for small $\Gamma_{\rm S}$ simply slows the approach to 
the asymptotic free-spin behavior at $T=0$ in singly-occupied 
regime ($|\delta| \ll U$). 
However, at $\Gamma_{\rm S}=\sqrt{U^2/4-\delta^2}$ there is a 
quantum phase transition from spin doublet for small $\Gamma_{\rm S}$ 
to spin singlet at large $\Gamma_{\rm S}$ \cite{Gorski2018Oct}, and $\mu_{\rm DS}(0)$ 
discontinuously switches from $1/4$ to $0$. 
This is in a stark contrast to continuous and always incomplete 
suppression of $\mu_{\rm DM}$, cf.~\eq{muDM}.

In the presence of a normal lead, the situation changes dramatically. 
Already for $\lambda=0$ (\ie{} with MW completely detached) $H$ takes 
the form of the Anderson model, where the Kondo effect leads to
screening of $\mu(T)$ below $T\approx T_K$ \cite{Hewson1997}.
However, $T_K$ rapidly decreases for small $\Gamma$, such that
in reality at lowest experimentally relevant temperature $T$ 
(modeled here by $T=10^{-7}U$)
a crossover is observed between the weak coupling regime for small $\Gamma$, 
with $\mu(T) \approx 1/4$, and the strong coupling regime for
large $\Gamma$, characterized by the Kondo-screened moment, $\mu(T)=0$.
The weak coupling case is realized \eg{} for $\Gamma=10^{-3}U$,
as presented in \fig{T} with a solid black line, lying on top 
of the dashed free-spin ($\lambda=0$) one.

While without MW $\Gamma=10^{-3}U$ is too small to affect the fate 
of the magnetic moment at relevant temperatures 
(corresponding $T_K \sim 10^{-172}U$ \cite{Haldane1978Feb}),
the presence of MW changes this situation dramatically. This is
clearly visible as a difference between dashed and solid lines 
in \fig{T} for $\lambda>0$. In all these cases $\mu(T)$ is
partially suppressed as for $\Gamma=0$ at intermediate $T$,
 but drops to $0$ for $T\to 0$, similarly to 
the Kondo regime. This is even more evident in \fig{Maj},
where $\mu$ is plotted as a function of $\lambda$ for a few
values of $\Gamma$ at $T=10^{-7}U$, mimicking a cryogenic 
experiment. Even for $\Gamma=10^{-5}U$, large $\lambda$
leads to complete suppression of $\mu$, which requires 
screening by conduction band electrons. This shows that
presence of MW vastly increases the Kondo temperature, 
by hasting the renormalization group flow away from the 
local moment fixed point at high energies, before the
Kondo coupling becomes relevant there. Apparently, the 
Kondo coupling is even more relevant around $\Gamma=0$
fixed point for $\lambda>0$. Not only does it scale to strong
coupling \cite{Cheng2014Sep,Vernek2019}, but also it does 
so at much higher $T$. This result agrees with general 
tendencies of $\lambda$ increasing $T_K$ reported in 
Refs.~\cite{Ruiz-Tijerina2015Mar,Weymann2017Apr}, but due
to large $\lambda$ considered here the effect is much more 
spectacular. This intriguing effect calls for a better 
understanding.

\section{Conclusions}
\label{sec:conclusions}

Our analytical results for a quantum dot coupled to large-gap 
topological superconductor wire in the absence of a normal 
leads show a universal partial suppression of the QD’s local 
magnetic moment for strong QD-MW coupling $\lambda$. This is in contrast 
to conventional superconductor behavior, where the low-temperature
magnetic moment does not change. In the presence of the normal 
lead, even smaller magnitude of $\lambda$ causes a tremendous increase 
of the Kondo temperature. In particular, $\lambda=0.01U$ is sufficient 
to enhance it for $\Gamma =0.001U$ from hundreds of orders of magnitude 
below $U$ to around $10^{-7}U$. Together with recent reinterpretation 
\cite{Svetogorov2023Jan} of existing experimental data concerning candidate Majorana 
observation \cite{Razmadze2020Sep} in terms of conventional Kondo effect, this shows 
that proper understanding of the Kondo physics in Majorana systems might 
be crucial for correct interpretation of the measurements.

\begin{center}
\textbf{Acknowledgments.}
\end{center}
\vspace{-2ex}
Work funded by Polish National Science Centre through 
grant no.~2018/29/B/ST3/00937 (KPW) and 2021/41/N/ST3/01885 (PM).
KPW acknowledges support from the Alexander von Humboldt Foundation.


%

\end{document}